\documentclass[10pt,notitlepage,nofootinbib,superscriptaddress,showpacs,showkeys,preprintnumbers,floatfix,byrevtex, twocolumn]{revtex4-2}

\usepackage{soul}										
\setstcolor{red}	                                              

\usepackage{epsfig}
\usepackage{amsmath,amssymb}
\usepackage{bm}
\usepackage{graphicx}
\usepackage{cancel}
\usepackage{empheq}
\usepackage{fancyvrb}

\usepackage{booktabs}
\usepackage{hyperref}
\usepackage{lineno}

\usepackage{color}
\usepackage{multirow}
\usepackage{stackengine}

\newcommand\xrowht[2][0]{\addstackgap[.5\dimexpr#2\relax]{\vphantom{#1}}}

\begin{document}

\preprint{Preprint}

\title{Effects of neutron-rich nuclei masses on symmetry energy}

\author{Seonghyun Kim}
\affiliation{Department of Physics and OMEG Institute, Soongsil University, Seoul 156-743, Republic of Korea}

\author{Dukjae Jang}%
\email{Corresponding Author: djjang2@ibs.re.kr}
\affiliation{Center for Relativistic Laser Science, Institute for Basic Science (IBS), Gwangju 61005, Republic of Korea}

\author{Soonchul Choi}
\email{Corresponding Author: scchoi0211@ibs.re.kr}
\affiliation{Center for Exotic Nuclear Studies, Institute for Basic Science (IBS), Daejeon 34126, Republic of Korea}

\author{Tsuyoshi Miyatsu}
\affiliation{Department of Physics and OMEG Institute, Soongsil University, Seoul 156-743, Republic of Korea}

\author{Myung-Ki Cheoun}
\affiliation{Department of Physics and OMEG Institute, Soongsil University, Seoul 156-743, Republic of Korea}

\date{\today}

\begin{abstract}
We explore the impact of neutron-rich nuclei masses on the symmetry energy properties using the mass table evaluated by the deformed relativistic Hartree-Bogoliubov theory in continuum (DRHBc) model. First, using the semi-empirical mass formula with the DRHBc mass table, we investigate the symmetry energy at saturation density $\rho_0$, denoted as $S_0$, and the ratio of surface to volume contributions to the symmetry energy, $\kappa$. As a result, we obtain $S_0=27.85\,{\rm MeV}$ ($\kappa=1.38$) for $a_{\rm sym}(A) =S_0 (1 - \kappa A^{-1/3})$ (Type I) and $S_0=32.66\,{\rm MeV}$ ($\kappa=3.15$) for $a_{\rm sym}(A) = S_0 (1 + \kappa A^{-1/3} )^{-1}$ (Type II), which are lower than those obtained using the AME2020 mass table, $S_0=28.54\,{\rm MeV}$ ($\kappa=1.29$) for Type I and $S_0=33.81\,{\rm MeV}$ ($\kappa=3.04$) for Type II. Second, we further investigate the effect of these changes in $a_{\rm sym}(A)$ on the density-dependent symmetry energy by employing the empirical model of $S(\rho) = C_k(\rho/\rho_0)^{2/3} + C_1(\rho/\rho_0) + C_2(\rho/\rho_0)^{\gamma}$ and universal relation of $a_{\rm sym}(A=208) = S(\rho=0.1\,{\rm fm}^{-3})$. Compared to the experimental constraints, we find that $S_0$ and slope parameter $L$, determined by the DRHBc mass table with Type II, are more suitable to explain the constraints by heavy ion collisions and isobaric analog states than AME2020. We also discuss the neutron skin thickness derived from the $L$, comparing it with experimental measurements.
\end{abstract}

\maketitle

\section{Introduction \label{S1}}
The nuclear symmetry energy plays an crucial role of understanding some experimental data of finite nuclei and lots of properties of isospin-asymmetric nuclear matter~\cite{Danielewicz:2008cm,Tsang:2012se,Horowitz:2014bja}.
Around the nuclear saturation density, $\rho_0$, the density-dependent symmetry energy is generally expanded as $S(\rho) \simeq S_0+L\chi + \mathcal{O}(\chi^2)$ with $\chi = (\rho-\rho_0)/(3\rho_0)$, where $S_0$ and $L$ denote the symmetry energy and the slope parameter at the $\rho_0$, respectively \cite{Baran:2004ih,Chen:2004si}.
The properties of the symmetry energy, including $S_0$ and $L$, can be determined from various measurements such as heavy-ion collisions (HICs) \cite{Famiano:2006rb,SRIT:2021gcy}, neutron skin thickness measurements via parity-violating elastic electron scattering \cite{Typel:2001lcw,Vinas:2013hua,Reed:2021nqk}, and astrophysical observations of neutron stars \cite{Steiner:2004fi,Lattimer:2006xb}. However, current determinations based on various experimental measurements still span a broad range of values, with $24\le S_0\,\mathrm{(MeV)}\le36$ and $-10\le L\,\mathrm{(MeV)}\le130$~\cite{Lattimer:2014sga,Tsang:2019mlz,Newton:2020jwn}, making it challenging to determine the precise values of $S_0$ and $L$~\cite{Piekarewicz:2021jte,Reinhard:2021utv}.

The symmetry energy coefficient of finite nuclei, $a_{\rm sym}(A)$, is also a key quantity to study their characteristics because it can be directly provided by nuclear masses which are the most precisely measured information in nuclear physics.
Using the semi-empirical mass formula, known as Bethe--Weizs\"{a}cker mass formula~\cite{Weizsacker:1935bkz,Bethe:1936zz}, $a_{\rm sym}(A)$ is extracted from the mass differences of isotope or isobaric nuclei~\cite{Wang:2010ra,Zhang:2013wna,Tian:2014uka}, the measured $\alpha$-decay energies of heavy nuclei~\cite{Dong:2010pw,Dong:2012ah}, and the double differences of ``experimental'' symmetry energies~\cite{Jiang:2012zzb}.
In particular, it has been proposed that a universal relation exists between $a_{\rm sym}(A)$ and $S(\rho)$ in mean-field theories, $a_{\rm sym}(A=208) \simeq S(\rho=0.1\,{\rm fm}^{-3})$ \cite{Centelles:2008vu}. This relation enables us to evaluate nuclear matter properties using information derived from finite nuclei, such as the neutron skin thickness and electric dipole polarizability (EDP) of $^{208}{\rm Pb}$ \cite{Centelles:2008vu,Zhang:2015ava}, which might be a key relation for further discussion.

\begin{figure*}[t]
\includegraphics[width=15 cm]{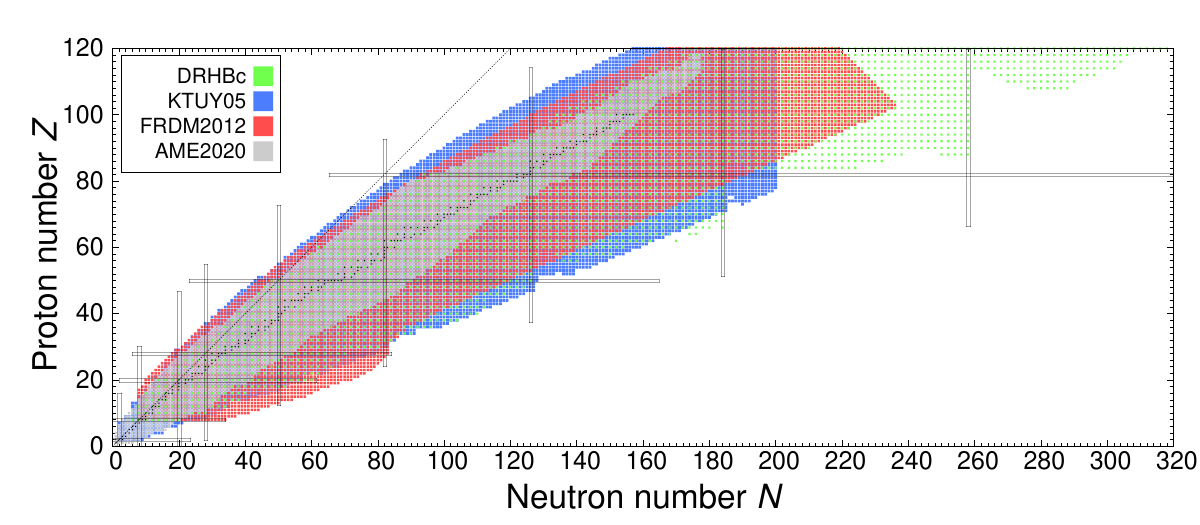}
\caption{Mass range covered by each mass table. Green, blue, red, and grey colored regions are covered by DRHBc \cite{DRHBcMassTable:2022uhi}, KTUY05 \cite{Koura:2000sle}, FRDM2012 \cite{Moller:2015fba}, and AME2020 \cite{Wang:2021xhn} mass tables, respectively.}
\label{fig:mass}
\end{figure*}

The extraction of the $a_{\text{sym}}(A)$ from the mass formula relies on the determination of nuclear binding energy. In the past decade, significant advancements have been made in the development of several nuclear mass tables. Notably, the KTUY05 model has introduced a mass formula incorporating shell energy corrections \cite{Koura:2000sle}, and a comprehensive evaluation of nuclear masses for 9318 nuclei has been constructed by using the finite-range droplet macroscopic (FRDM) and the folded-Yukawa single-particle microscopic mass models (FRDM2012) \cite{Moller:2015fba}. Moreover, the atomic mass evaluations, AME2020, have provided nuclear mass data for 2550 stable nuclei in their ground states, based on experimentally measured nuclear masses \cite{Wang:2021xhn}. Recent efforts have been directed towards expanding the nuclear mass table to include the neutron drip line, employing the deformed relativistic Hartree-Bogoliubov theory in continuum (DRHBc) model. This extends mass table encompasses 2583 even-even nuclei, spanning from the proton drip line to the neutron drip line \cite{DRHBcMassTable:2022uhi}. Fig.\,\ref{fig:mass} depicts the coverage of each mass table.

In this study, we explore the impact of neutron-rich nuclei on $a_{\rm sym}(A)$ by adopting the DRHBc and AME2020 mass tables.
As shown in Fig.\,\ref{fig:mass}, the DRHBc mass table provides a broader coverage of nuclear masses, extending to neutron-rich nuclei, compared to the AME2020 mass table which is limited to the experimental data region. We show how the nuclear masses of neutron-rich nuclei in DRHBc mass table affect the $a_{\rm sym}(A)$. Furthermore, we present implications of the change in $a_{\rm sym}(A)$ for $S(\rho)$ by employing the universal relation $S(\rho=0.1\,{\rm fm}^{-3}) = a_{\rm sym}(A=208)$ \cite{Centelles:2008vu} and compare the results with experimental constraints from heavy-ion collisions, measurements in finite nuclei, and observations of neutron stars.

This paper is organized as follows.
In Sec.~\ref{S2}, we present $a_{\rm sym}(A)$ with the DRHBc and AME2020 mass tables.
In Sec.~\ref{S3}, we discuss the effects of change in $a_{\rm sym}(A)$ on $S(\rho)$.
Lastly, a summary is included in Sec.~\ref{S4}.

\section{The symmetry energy coefficient with neutron-rich nuclei \label{S2}}
In the Bethe--Weizs\"{a}cker mass formula, the binding energy of a nucleus with mass number $A (=N+Z)$ is given by
\begin{align}
  B(A,Z) &= a_v A-a_{\rm surf} A^{2/3} \nonumber \\
         &- a_{\rm sym}\frac{(Z-N)^2}{A} - E_{\rm Coul}(A,Z) +a_{\rm pair} A^{-1/2},
  \label{eq:BWmass}
\end{align}
where $a_{v(\mathrm{surf})[\mathrm{pair}]}$ stands for the coefficient of volume (surface) [pairing] term and $E_{\rm Coul}$ is the Coulomb energy.
Taking into account the difference of binding energies between isobaric nuclei, the symmetry energy coefficient of finite nuclei, $a_{\rm sym}(A,Z,n)$, is written as
\begin{align}
  a_{\rm sym}(A,Z,n)
  &= \frac{A}{8(A-2Z)}\Biggl[\frac{B(A,Z+n)-B(A,Z-n)}{n} \nonumber \\
  &\qquad - \frac{E_{\rm Coul}(A,Z+n)-E_{\rm Coul}(A,Z-n)}{n}\Biggr],
  \label{eq:deri-asym}
\end{align}
with $n$ being a positive integer that determines the binding energy difference of isobaric nuclei.
Although the general form of $a_{\rm sym}$ is expressed as a function of $A$, we explicitly present $Z$ as well as $A$ in Eq.\,\eqref{eq:deri-asym} to figure out the isospin dependence on $a_{\rm sym}$ using the mass table with neutron-rich nuclei.
To avoid the choice of a reference nucleus used in Ref.\,\cite{Tian:2014uka}, we simply consider the mean value of Eq.\,\eqref{eq:deri-asym}:
\begin{equation}
  \tilde{a}_{\rm sym}(A,Z) = \frac{1}{m}\sum_{n=1}^{m}a_{\rm sym}(A,Z,n),
  \label{eq:asym_avg}
\end{equation}
with $m$ being the number of pairs of isobaric nuclei.
In addition, we take the average of Eq.\,\eqref{eq:asym_avg} to compare it with the conventional symmetry energy coefficient of finite nuclei only with the $A$ dependence:
\begin{equation}
   \bar{a}_{\rm sym}(A)= \frac{1}{ k }\sum_{Z={Z}_{\rm min}}^{{Z}_{\rm max}}\tilde{a}_{\rm sym}(A,Z),
  \label{eq:asym_2avg}
\end{equation}
where ${Z}_{\rm max(min)}$ denotes the maximum (minimum) number of $Z$ and $k$ is the number of $\tilde{a}_{\rm sym}(A,Z)$ in a given isobaric chain. Hereafter, we use $a_{\rm sym}(A)$ for $\bar{a}_{\rm sym}(A)$.

In this study, we employ two phenomenological functions $a_{\rm sym}(A)$ to fit the data obtained by Eq.\,\eqref{eq:asym_2avg} from the given mass tables.
One is $a_{\rm sym}(A)=S_0(1-\kappa A^{-1/3})$ (Type I) and the other is given by $a_{\rm sym}(A)=S_0(1+\kappa A^{-1/3})^{-1}$ (Type II) with the parameters $S_0$ and $\kappa$, where $\kappa$ indicates the ratio of surface to volume contributions of the $a_{\rm sym}(A)$, i.e., $\kappa = a_{\rm sym}^S(A)/a_{\rm sym}^V(A)$ \cite{Tian:2014uka}.
We can see that Type I corresponds to the first order expansion of Type II in the small limit of $A$.
In both forms, the $S_0$ is dominant in the large $A$, while the $\kappa$ becomes effective in the small $A$.

To precisely evaluate the $a_{\rm sym}(A)$, it is necessary to remove the microscopic shell corrections from their binding energies because those corrections are not considered in the Bethe--Weizs\"{a}cker mass formula.
This is the same as in the case of Wigner correction.
The binding energy of a nucleus in Eq.~\eqref{eq:BWmass} is hence given by $B(A,Z) = B_{\rm Data}(A,Z)-E_{\rm sh}(A,Z)-E_{W}(A,Z)$, where $B_{\rm Data}(A,Z)$ is the experimental data taken from the AME2020 or DRHBc mass tables, $E_{\rm sh}(A,Z)$ is the shell correction energy, and $E_{W}(A,Z)$ is the Wigner correction~\cite{Tian:2014uka}.
We here adopt $E_{\rm sh}(A,Z)$ from the KTUY05 mass formula~\cite{Koura:2000sle} since the shell corrections in the DRHBc mass table have not been studied yet.
We also use the form of $E_W(A,Z) = 10 \exp (-4.2 |I|)$ with isospin asymmetry, $I=(N-Z)/A$~\cite{Liu:2010ne}.

As for the Coulomb energy in Eq.~\eqref{eq:BWmass}, we exploit the same expression in Ref.~\cite{Tian:2014uka}, deduced from the 88 pairs of mirror nuclei in the region of $11 \le A \le 75$: $E_{\rm Coul}(A,Z) = a_{\rm Coul} Z(Z-1)(1-bZ^{-2/3})/A^{1/3}$ with $a_{\rm Coul}=0.704$ MeV and $b=0.985$.

\begin{figure}[t]
  \centering
  \includegraphics[width=7.5cm]{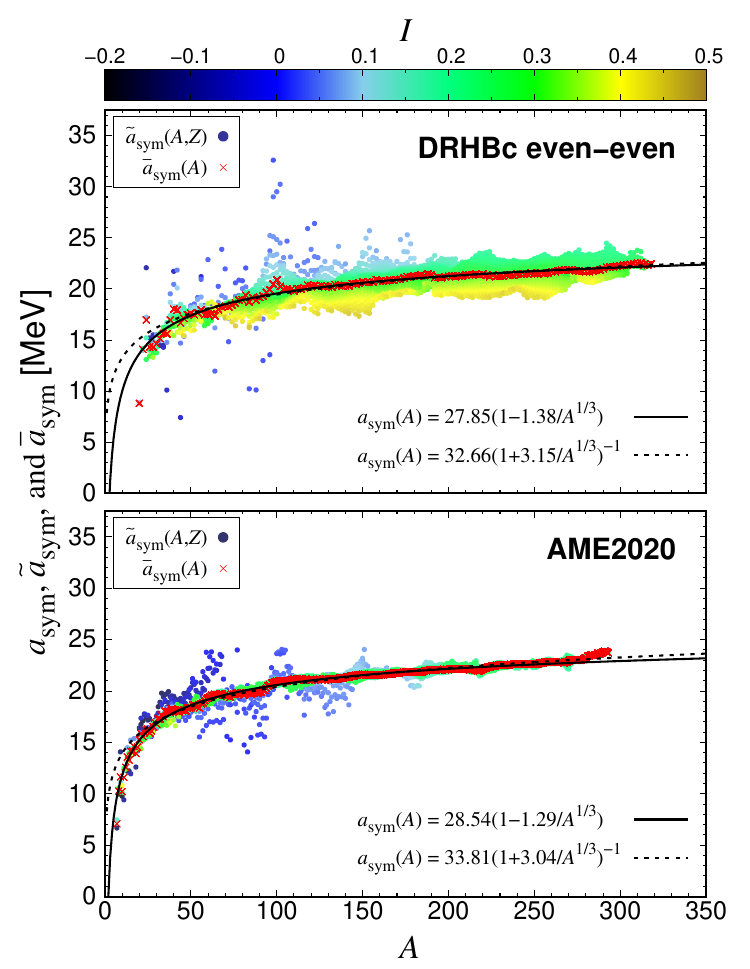}
  \caption{
    The $\tilde{a}_{\rm sym}(A,Z)$ in Eq.~\eqref{eq:asym_avg} (colored points) and $\bar{a}_{\rm sym}(A)$ in Eq.~\eqref{eq:asym_2avg} (red cross mark). The upper (lower) panel is for the case of DRHBc (AME2020) mass table. The solid (dashed) lines are the best-fitting curves of $a_{\rm sym}(A)$ with Types I (II) in both panels. The color bar indicates the values of isospin asymmetry, $I \equiv (N-Z)/A$.}
  \label{fig1}
\end{figure}
Fig.\,\ref{fig1} shows the symmetry energy coefficients of finite nuclei, $\tilde{a}_{\rm sym}(A,Z)$ in Eq.~\eqref{eq:asym_avg} and $\bar{a}_{\rm sym}(A)$ in Eq.~\eqref{eq:asym_2avg}, with the mass tables of DRHBc (upper panel) and AME2020 (lower panel). The extracted $\tilde{a}_{\rm sym}(A,Z)$ from the DRHBc mass table extensively covers the neutron-rich nuclei region compared to that of the AME2020 mass table. In particular, in Fig.~\ref{fig1}, $\tilde{a}_{\rm sym}(A)$s for the neutron rich nuclei (yellow points) suppress $\bar{a}_{\rm sym}(A)$, resulting in reduction of $S_0$. As a result, we obtain the $S_0=27.85$ MeV (Type I) and $S_0=32.66$ MeV (Type II) with the DRHBc mass table. On the other hand, for the AME2020 mass table, we obtain $S_0=28.54$ MeV (Type I) and $S_0=33.81$ MeV (Type II). This implies that the binding energies of neutron-rich nuclei contribute to a reduction in $S_0$. Furthermore, it is noteworthy that a more substantial decrease in $S_0$ would occur if a broader range of neutron-rich nuclides could be considered. However, the inclusion of such nuclides in the DRHBc mass table was limited by the availability of shell correction data adopted from KTUY05 data.

\section{Effects of changes in $a_{\rm sym}(A)$ on nuclear matter properties}
\label{S3}
We evaluate the effects of the change in $a_{\rm sym}(A)$ due to the neutron-rich nuclei on $S(\rho)$ by employing the following empirical density-dependent symmetry energy model \cite{Dong:2012zza, Dong:2012ah}: 
\begin{eqnarray}
S(\rho) = C_k (\rho/\rho_0)^{2/3} + C_1 (\rho/\rho_0) + C_2 (\rho/\rho_0)^{\gamma}.
\label{eq:Srho}
\end{eqnarray}
We take the previous determinations of $C_k$ and $\gamma$ from the correlations in symmetry energy parameters, $C_k=17.47\,{\rm MeV}$ and $\gamma=1.52$ \cite{Dong:2012zza}. In addition, to determine the remained coefficients $C_1$ and $C_2$, we adopt two relations of $S(\rho =\rho_0) = S_0$ and $S(\rho=0.1\,{\rm fm}^{-3}) \simeq a_{\rm sym}(A=208)$ \cite{Centelles:2008vu}. We note that the DRHBc (AME2020) mass table results in $a_{\rm sym}(A=208) = 21.36\,{\rm MeV}$ ($a_{\rm sym}(A=208) = 22.31\,{\rm MeV}$) for Type I and $a_{\rm sym}(A=208) = 21.32\,{\rm MeV}$ ($a_{\rm sym}(A=208) = 22.33\,{\rm MeV}$) for Type II. For $\rho_0$, we adopt $\rho_0 = 0.15 \pm 0.01 \,{\rm fm}^{-3}$ \cite{Horowitz:2020evx}. Taking into account the two conditions, we determine $C_1$ and $C_2$ for each result from the DRHBc and AME2020 mass tables. Moreover, using the relation of $L = 2C_k + 3C_1 + 3C_2\gamma$, we evaluate the $L$. We tabulate determinations of $C_1$, $C_2$, $S_0$, and $L$ in Tab.\,\ref{table1}, in which upper and lower limits for each data stem from the uncertainty in $\rho_0$.
\begin{table}[t]
  \caption{Determinations of $C_1$, $C_2$, $S_0$, and $L$ for Types I and II with DRHBc and AME2020 mass tables. The uncertainties stem from that of $\rho_0$, i.e., $\rho_0 = 0.15 \pm 0.01\,{\rm fm^{-3}}$.}
     \label{table1}
      \centering
         \begin{tabular}{c c|c|c|c|c||c|c|c|c|}
           \cline{1-3} \cline{3-6} \cline{7-10}  
           \multirow{3}{*}{ } & \multirow{3}{*}{ } & \multicolumn{4}{c||}{DRHBc}&\multicolumn{4}{c|}{AME2020} \\
           \cline{3-6} \cline{7-10} 
            & & \multicolumn{2}{c|}{Type I} & \multicolumn{2}{c||}{Type II} & \multicolumn{2}{c|}{Type I}  & \multicolumn{2}{c|}{Type II}  \\
            \hline\xrowht[()]{11pt}
            \multirow{2}{*}{} & $C_1\,[{\rm MeV}]$ & \multicolumn{2}{c|}{$19.1^{+6.8}_{-8.9}$} & \multicolumn{2}{c||}{$-1.6^{+9.9}_{-13.6}$} & \multicolumn{2}{c|}{$23.7^{+6.7}_{-8.7}$} & \multicolumn{2}{c|}{$1.4^{+10.1}_{-13.8}$} \\ 
            \hline\xrowht[()]{11pt}
            &$C_2\,[{\rm MeV}]$& \multicolumn{2}{c|}{$-8.7^{+8.9}_{-6.8}$} & \multicolumn{2}{c||}{$16.8^{+13.6}_{-9.9}$} & \multicolumn{2}{c|}{$-12.7^{+8.7}_{-6.7}$} & \multicolumn{2}{c|}{$14.9^{+13.8}_{-10.1}$}\\ [0.5ex]
            \hline\xrowht[()]{11pt}
            &$S_0\,[{\rm MeV}]$& \multicolumn{2}{c|}{$27.85$} & \multicolumn{2}{c||}{$32.66$} & \multicolumn{2}{c|}{$28.54$} & \multicolumn{2}{c|}{$33.81$}\\ [0.5ex]
            \hline\xrowht[()]{11pt}
            &$L\,[{\rm MeV}]$& \multicolumn{2}{c|}{$52.5^{+13.8}_{-10.6}$} & \multicolumn{2}{c||}{$106.7^{+21.1}_{-15.5}$} & \multicolumn{2}{c|}{$48.4^{+13.6}_{-10.5}$} & \multicolumn{2}{c|}{$107.2^{+21.6}_{-15.8}$}\\ [0.5ex]
            \hline\xrowht[()]{11pt}
         \end{tabular}
\end{table}

\begin{figure}[t]
\centering
\includegraphics[width=7.5cm]{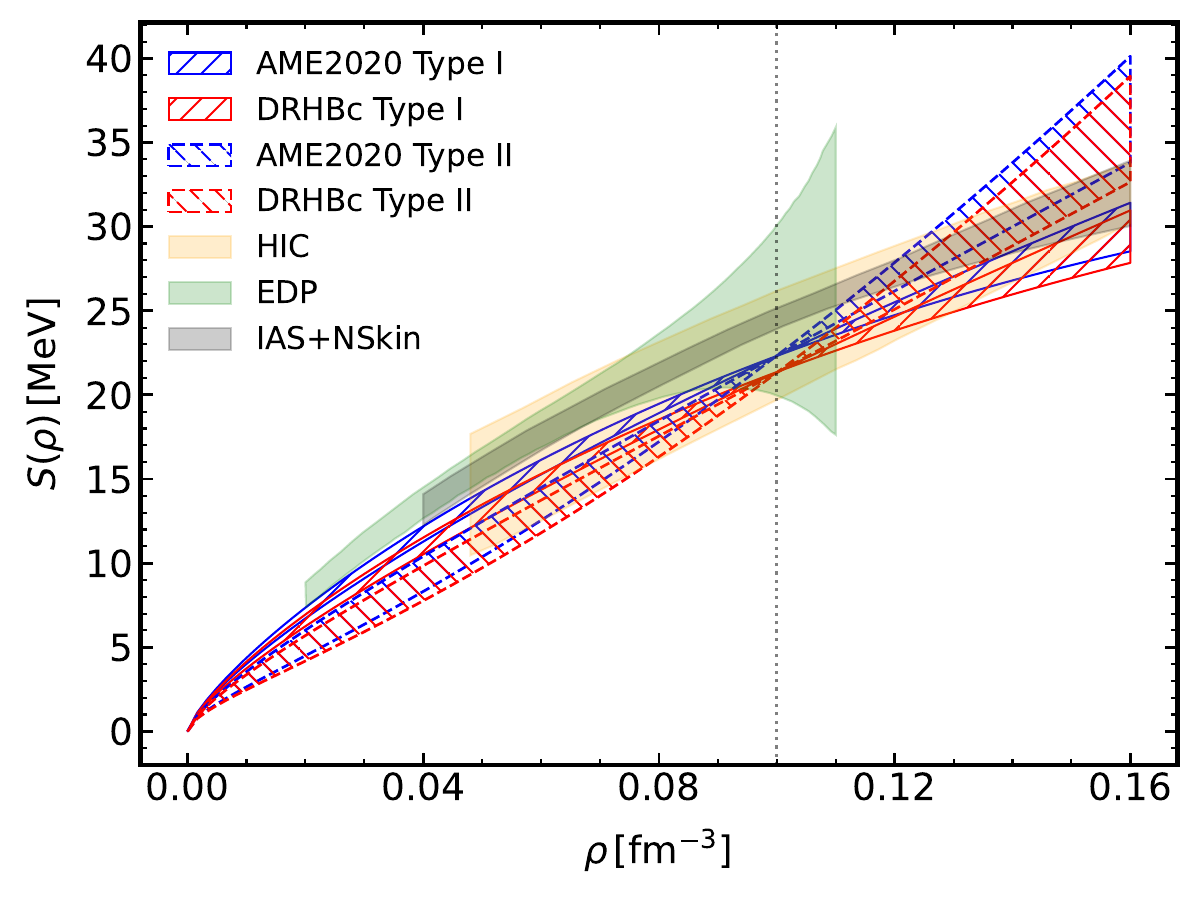}
\caption{$S(\rho) = C_k(\rho/\rho_0)^{2/3} + C_1(\rho/\rho_0) + C_2(\rho/\rho_0)^\gamma$ as a function of $\rho$. The hatches with (off-)diagonal lines indicate Type I (Type II) regions. In each Type, the red (blue) colored region corresponds to the results obtained using the DRHBc (AME2020) mass table. The orange, green, and grey colored regions represent the experimental constraints from analyses of HIC \cite{Tsang:2008fd}, EDP measurement in $^{208}{\rm Pb}$ \cite{Zhang:2015ava}, and IAS+Nskin data \cite{Danielewicz:2013upa}, respectively. }
\label{Esym}
\end{figure}

Fig.\,\ref{Esym} shows the evaluated $S(\rho)$ from the determinations in Tab.\,\ref{table1} as a function of $\rho$ with experimental constraints from analyses of EDP in $^{208}{\rm Pb}$ \cite{Zhang:2015ava}, HICs \cite{Tsang:2008fd}, and the isobaric analog states with neutron skin (IAS+Nskin) data \cite{Danielewicz:2013upa}. The EDP measurement provides constraints on $S(\rho)$ at $\rho \lesssim 2\rho_0/3$, which are consistent with our determinations of $S(\rho=0.1\,{\rm fm}^{-3})$ for both types regardless of the mass table. On the other hand, the constraints of $S(\rho)$ around $\rho_0$ are provided from the analyses of HIC and IAS+Nskin, in which the allowed range of the $S(\rho)$ depends on the uncertainty of evaluated $S(\rho)$.

The behavior of $S(\rho)$ depends on three conditions. First, the condition of $S_0 = S(\rho = \rho_0)$ determines the behavior of $S(\rho)$ in the vicinity of $\rho \approx \rho_0$. Since an increase in $S_0$ leads to a higher value of $S(\rho=\rho_0)$, $S(\rho)$ becomes stiffer as $S_0$ increases for the fixed $S(\rho \simeq 0.1\,{\rm fm}^{-3})$ \cite{Dong:2012ah}. Therefore, the $S(\rho)$ for Type II ($S_0 \simeq 33\,{\rm MeV}$) is stiffer than the Type I ($S_0\simeq 28\,{\rm MeV}$), which results in the higher $L$ for Type II than that of Type I. (See Tab.\,\ref{table1}.)

Second, the stiffness of $S(\rho)$ depends on the condition of not only $S_0$, but $S(\rho = 0.1\,{\rm fm}^{-3})$. In Fig.\,\ref{Esym}, there are two intersecting points at $\rho=0.1\,{\rm fm}^{-3}$. Each point stems from the condition of $S(\rho = 0.1\,{\rm fm}^{-3}) = a_{\rm sym}(A=208)$ for each mass table. Since the fitted line for the DRHBc mass table in Fig.\,\ref{fig1} leads to a lower value of $a_{\text{sym}}(A=208)$ compared to the AME2020 mass table, the intersecting point for the DRHBc mass table in Fig.\,\ref{Esym} is lower than the case of AME2020 mass table. This lower value of $S(\rho=0.1\,{\rm fm}^{-3})$ contributes to make a stiffer $S(\rho)$. As a result, when we compare the $S(\rho)$ for each mass table in Type I, $S(\rho)$ with DRHBc mass table is stiffer than that of the AME2020 case, despite its lower $S_0$. However, in the case of Type II, the difference of $S_0$ between DRHBc and AME2020 is greater than that of Type I, so that the $S(\rho)$ for AME2020 is slightly stiffer than the $S(\rho)$ for the DRHBc mass table. Consequently, in Tab.\,\ref{table1}, the $L$ for DRHBc in Type I (Type II) is higher (lower) than the $L$ for AME2020.

Third, the $L$ depends on $\rho_0$. For the given two conditions of $S(\rho=0.1\,{\rm fm}^{-3})=a_{\rm sym}(A=208)$ and $S(\rho=\rho_0) = S_0$, $S(\rho)$ in Eq.\,(\ref{eq:Srho}) decreases, as $\rho_0$ increases. In this case, the $S(\rho)$ becomes softer, which in turn reduces $L$, and vice versa. This is shown in Tab.\,\ref{table1}, where the upper (lower) limits of $L$ correspond to the results with lower (upper) limit of $\rho_0$, respectively.

\begin{figure}[t]
\centering
\includegraphics[width=7.5cm]{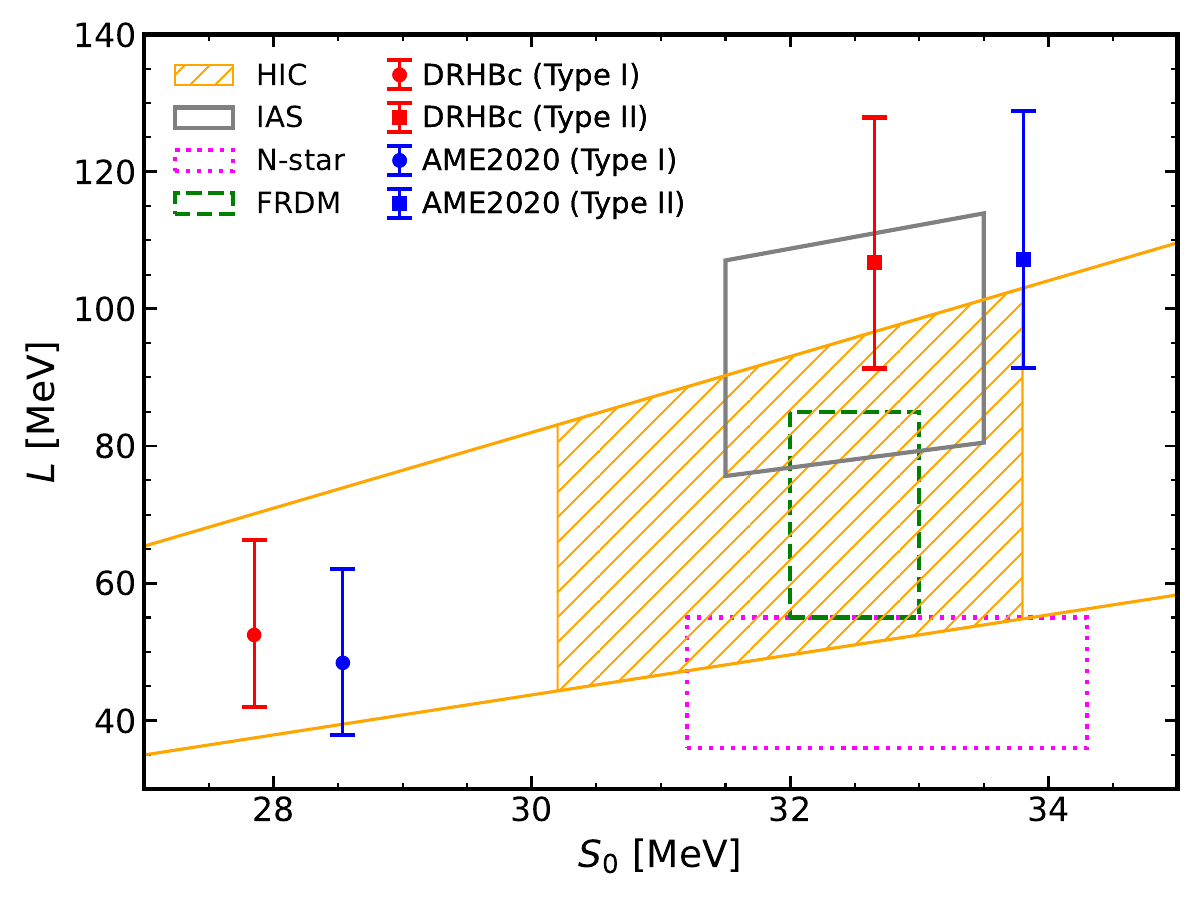}
\caption{Determinations of $S_0$ and $L$ with various constraints. The red (blue) line with circle symbols represents the determination of $S_0$ and $L$ for DRHBc (AME2020) Type I. The red (blue) line with square symbols indicates the determination of $S_0$ and $L$ for DRHBc (AME2020) Type II. In each line, the upper and lower limits stem from the uncertainty in $\rho_0$, while the circle and square symbols correspond to results at $\rho_0=0.15\,{\rm fm^{-3}}$. The orange-colored lines show the constraints from HICs \cite{Tsang:2008fd}, and the region with hatched diagonal lines correspond to the region including the pygmy dipole resonance data \cite{Klimkiewicz:2007zz}. The gray-solid, green-dashed, and magenta-dotted boxes denote the constraints from the IAS \cite{Danielewicz:2008cm}, the FRDM \cite{Moller:2012pxr}, and observations of neutron stars (labeled by `N-star') \cite{Steiner:2011ft}, respectively.}
\label{L}
\end{figure}

We compare our determinations of $S_0$ and $L$ with experimental and observational constraints in Fig.\,\ref{L}. The orange-colored lines represent the constrained region of $S_0$ and $L$ from the HICs experiments involving collisions between $^{112}{\rm Sn}$ and $^{124}{\rm Sn}$ \cite{Tsang:2008fd}. Here, the region with hatched diagonal lines includes constraints from the pygmy dipole resonance data, yielding $30.2 \le S_0 ({\rm MeV}) \le 33.8$ \cite{Klimkiewicz:2007zz}. Consequently, out of the four cases considered in our determinations, only the $S_0$ value for Type II with the DRHBc mass table is allowed by this constraint. Furthermore, this region constrains the $L$ as $L \le 96.7\,{\rm MeV}$ for the DRHBc Type II, whose limit constrains $\rho_0$ as $\rho_0 \ge 0.156\,{\rm fm^{-3}}$. 

Measurements from finite nuclei also provide constraints on $S(\rho)$. The FRDM is advantageous to extract the symmetry energy from measured binding energies because it can precisely evaluate the contribution of each term in the empirical mass formula. We show the constraints from the FRDM by using the green dashed box in Fig.\,\ref{L}, which provides constraints on $S_0 = 32.5 \pm 0.5\,{\rm MeV}$ and $L = 70 \pm 15\,{\rm MeV}$ \cite{Moller:2012pxr}. The constraint on $S_0$ only allows the case of Type II with the DRHBc mass table. However, the constraints on $L$ excludes our determinations of $L$ for Type II. We also show constraints from the analysis of IAS \cite{Danielewicz:2008cm} by using the gray solid box in Fig.\,\ref{L}. Notably, this constraint only allows the determinations of $S_0$ and $L$ for the DRHBc Type II. In this case, $\rho_0$ is constrained as $\rho_0 \ge 0.148 \,{\rm fm}^{-3}$ by the given $L$.

Astrophysical observations are also one of the important constraints on the $S(\rho)$. For instance, the Quantum Monte Carlo (QMC) technique, effective approach to solve the many-body problem, has been combined with constraints on the mass and radius of neutron stars, which provides constraints of $31.2 < S_0 ({\rm MeV}) < 34.3$ and $36 < L ({\rm MeV}) < 55$ \cite{Steiner:2011ft}. We represent this constraint by using the magenta dotted line in Fig.\,\ref{L}. Our determinations of $S_0$ for Type II with both mass tables are allowed within this constraint, but the $L$ is excluded by the constraint. On the other hand, the values of $L$ for Type I are allowed, but the $S_0$ is excluded. We note that such astrophysical constraints also depend on uncertainties related to the description of X-ray bursts dynamics and the emissivity of the stellar surface \cite{Steiner:2010fz}. Therefore, there could exist a discrepancy between astronomical constraints and experimental constraints, which is expected to be improved with greater precision in the future.

\begin{figure}[t]
\centering
\includegraphics[width=7.5cm]{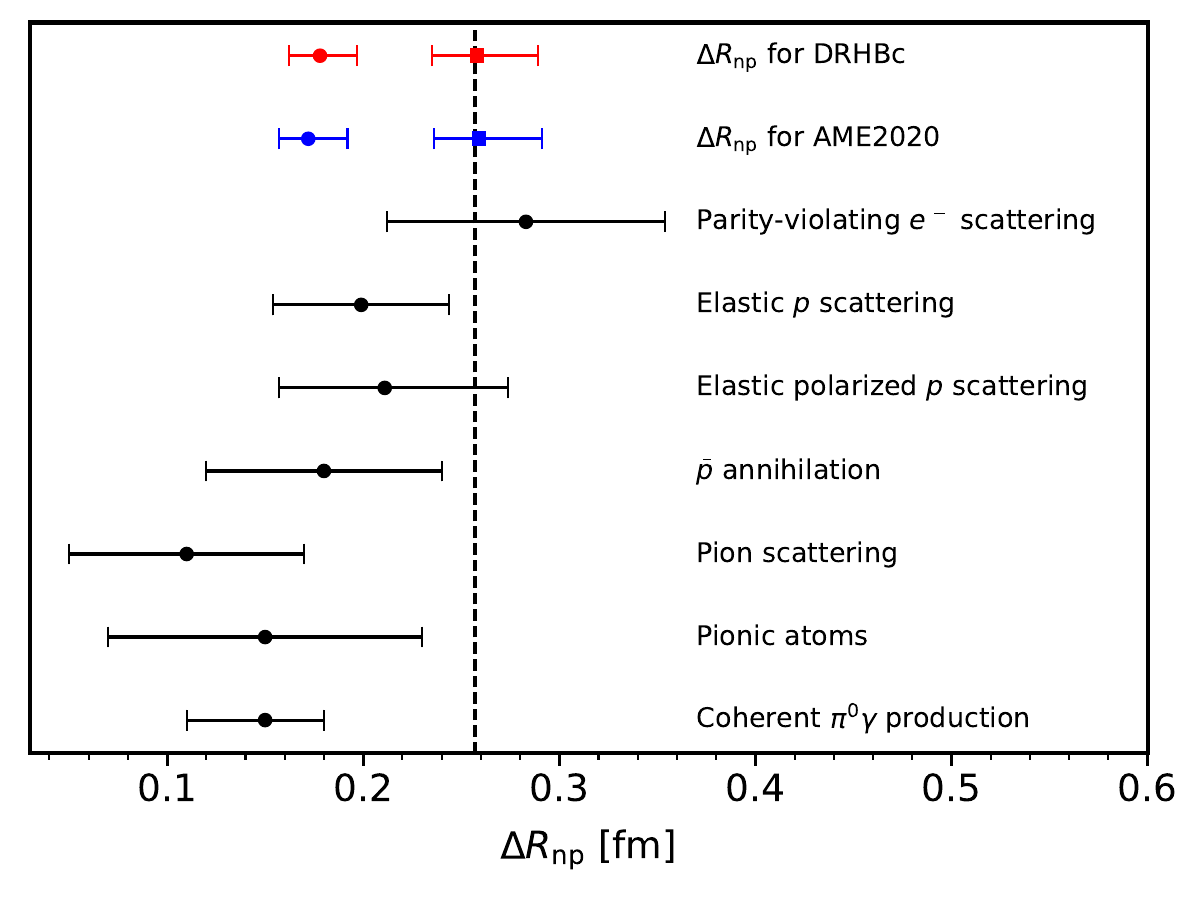}
\caption{Determinations of $\Delta R_{np}$ with experimental measurements. The red (blue) line with circle symbols represents $\Delta R_{np}$ for DRHBc (AME2020) Type I, and the line with square symbols represents $\Delta R_{np}$ for DRHBc (AME2020) Type II. The black lines with circle symbols denote the experimental measurements of $\Delta R_{np}$. For experimental measurements, we adopt the measured $\Delta R_{np}$ by including coherent $\pi^0\gamma$ production \cite{Tarbert:2013jze}, pionic atoms \cite{Friedman:2012pa}, $\pi$ scattering \cite{Friedman:2012pa}, $\bar{p}$ annihilation \cite{Brown:2007zzc,Klos:2007is},  elastic proton scattering \cite{Zenihiro:2010zz}, elastic polarized proton scattering \cite{Clark:2002se}, and Parity-violating $e^-$ scattering in PREX-2 \cite{PREX:2021umo}. The vertical black-dashed line indicates  $\Delta R_{\rm np} = 0.257\,{\rm fm}$, which is obtained from the microscopic calculations based on the DRHBc model \cite{Kim:2021skf}.}
\label{skin}
\end{figure}

Lastly, we discuss the effects of change in $a_{\rm sym}(A)$ on the neutron skin thickness, $\Delta R_{\rm np}$. Over the past decades, various methods have been employed to measure the $\Delta R_{\rm np}$, including coherent $\pi^0\gamma$ production \cite{Tarbert:2013jze}, pionic atoms \cite{Friedman:2012pa}, $\pi$ scattering \cite{Friedman:2012pa}, $\bar{p}$ annihilation \cite{Brown:2007zzc,Klos:2007is},  and elastic (polarized) proton scattering \cite{Ray:1979qv, Zenihiro:2010zz, Starodubsky:1994xt,Clark:2002se}. Recently, the PREX-2 collaboration reported new measurement of $\Delta R_{\rm np} = 0.283 \pm 0.071\,{\rm fm}$, using parity-violating electron scattering  \cite{PREX:2021umo}. To compare our determinations with those experimental measurements, we employ the relation of $\Delta R_{\rm np} = 0.101 + 0.00147 L$ \cite{Roca-Maza:2012uor}. As a result, for Type I, we obtain $\Delta R_{\rm np} = 0.178^{+0.019}_{-0.016}\,{\rm fm}$ ($0.172^{+0.020}_{-0.015}\,{\rm fm}$) from DRHBc (AME2020) mass table. For Type II, we obtain $\Delta R_{\rm np} = 0.258^{+0.031}_{-0.023}\,{\rm fm}$ ($0.259^{+0.032}_{-0.023}\,{\rm fm}$) from DRHBc (AME2020) mass table. These results are presented in Fig.\,\ref{skin} with other experimental determinations, in which the $\Delta R_{np}$ for Type II case is in agreement with the recent measurements from PREX-2. We also note that the $\Delta R_{np}$ for Type II is consistent with previous microscopic calculations based on the same DRHBc model, $\Delta R_{\rm np} = 0.257\,{\rm fm}$ \cite{Kim:2021skf}. Such self-consistency for $R_{np}$ between microscopic and macroscopic results could be a signal guaranteeing the present approach.

\section{Summary}
\label{S4}
In summary, we investigate the impact of neutron-rich nuclei masses on the properties of the symmetry energy using the DRHBc mass table. We find that the binding energies of neutron-rich nuclei can suppress $\bar{a}_{\rm sym}(A)$, resulting in decreased $S_0$. Specifically, we obtain $S_0=27.85\,{\rm MeV}$ ($\kappa=1.38$) for Type I and $S_0=32.66\,{\rm MeV}$ ($\kappa=3.15$) for Type II. These results of $S_0$ are reduced rather than the determinations from the AME2020 mass table, $S_0=28.54\,{\rm MeV}$ ($\kappa=1.29$) for Type I and $S_0=33.81\,{\rm MeV}$ ($\kappa=3.04$) for Type II. Furthermore, based on these results with the empirical form of $S(\rho) = C_k (\rho/\rho_0)^{2/3} + C_1 (\rho/\rho_0) + C_2(\rho/\rho_0)^{\gamma}$ and the two presumed conditions: $a_{\rm sym}(A=208) = S(\rho=0.1\,{\rm fm}^{-3})$ and $S_0 = S(\rho=\rho_0)$, we study properties of $S(\rho)$, $L$, and $\Delta R_{np}$ using the mass table results. We present a summary of all of the determinations in Tab.\,\ref{table1}.

Our findings reveal that changes in $\bar{a}_{\rm sym}(A)$ and $\rho_0$ affect the behavior of $S(\rho)$ under the assumption of the universal relation. Specifically, the results from the DRHBc (AME2020) mass table lead to a stiffer $S(\rho)$ for Type I (II), compared to the case of AME2020 (DRHBc) mass table. Interestingly, in the case of Type II, the decrease in $S_0$ due to the DRHBc mass table enables the determinations of $S_0$ to be allowed within the constraints from HICs and the IAS. In addition, the $L$ for this case is simultaneously allowed by these constraints depending on $\rho_0$. For each constraint on $L$, we provide the new constraints of $\rho_0$, $\rho_0 \ge 0.156\,{\rm fm}^{-3}$ for HICs and $\rho_0 \ge 0.148\,{\rm fm}^{-3}$ for IAS. Furthermore, we discuss the effects of change in $a_{\rm sym}(A)$ on the $\Delta R_{np}$. Notably, our evaluation of $\Delta R_{np}$ in Type II is consistent with previous microscopic calculation based on the DRHBc model as well as PREX-2 measurement.

These results presented in this study may change when considering more neutron-rich nuclei. Therefore, it is desirable to investigate the effects of contributions from additional neutron-rich nuclei near the neutron drip line on the $a_{\rm sym}(A)$. This study should involve a wider range of shell and Wigner corrections for the neutron-rich nuclei, which are not included in the current study. Such a future study will provide a more comprehensive understanding of how neutron-rich nuclei impact the properties of the symmetry energy.

\begin{acknowledgments}
S.K., T.M. and M.K.C. are supported by the National Research Foundation of Korea (Grant Nos. NRF-2020R1A2C3006177 and NRF-2021M7A1A1075764).
D.J. and S.C. are supported by the Institute for Basic Science under IBS-R012-D1 and IBS-R031-D1, respectively.
\end{acknowledgments}




\bibliography{./bib/Refs}

\end{document}